\documentstyle[epsf]{article}
 
\newcommand{\bee}{\begin{equation}}
\newcommand{\ee}{\end{equation}}
\newcommand{\beea}{\begin{eqnarray}}
\newcommand{\eea}{\end{eqnarray}}

\begin{document}
\thispagestyle{empty}
\parskip=12pt
\raggedbottom
 
\def\mytoday#1{{ } \ifcase\month \or
 January\or February\or March\or April\or May\or June\or
 July\or August\or September\or October\or November\or December\fi
 \space \number\year}
\noindent
\hspace*{9cm} COLO-HEP-420\\
\vspace*{1cm}
\begin{center}
{\LARGE The correlation length of the Heisenberg antiferromagnet
with arbitrary spin S}
 
\vspace{1cm}
Peter Hasenfratz\footnote{On leave from the Institute of Theoretical
Physics, University of Bern, Sidlerstrasse 5, 
\linebreak CH-3012 Bern, Switzerland.}
\\
Department of Physics\\
University of Colorado at Boulder \\
Boulder, CO 80309-0390

\vspace{0.5cm}
\mytoday \time \\ \vspace*{0.5cm}

\nopagebreak[4]
 
\begin{abstract}
Experiments and numerical data on the correlation length $\xi(T)$
for large $S$ disagree
strongly with the theoretical prediction based on the effective field
theory prescription of the magnon physics.
The reason is that for large $S$,
at any accessible $\xi(T)$, the cut-off effects from the non-magnon scales
become large and can not be treated by an effective field theory.
We study these effects in a spin-wave expansion.
The corrected prediction  on $\xi(T)$ connects the renormalized 
classical and the
classical scaling regions smoothly and comes close to the data.
\end{abstract}
 
\end{center}
\eject

There exist a number of quasi-2D antiferromagnetic compounds, including
spin-1/2 antiferromagnets (AFMs) ${\rm La_2CuO_4}$ and ${\rm Sr_2CuO_2Cl_2}$
\cite{S12EXP}, spin-1 AFMs ${\rm La_2NiO_4}$ and 
${\rm K_2NiF_4}$\cite{S10EXP} 
and spin-5/2 AFM ${\rm Rb_2MnF_4}$ 
\cite{S52EXP1,S52EXP2,S52EXP3} whose magnetic behavior is well 
described by the 2D quantum Heisenberg model\footnote{We use
$\hbar=1$, $k_{\rm B}=1$ convention.} 
\begin{equation}
\label{heis}
H = J \sum_{n,i} {\bf S}_{n+\hat{\imath}}{\bf S}_n \,, \qquad 
{\bf S}_n^2=S(S+1)\,.
\end{equation}
The low temperature properties of this model are dominated by magnon
excitations and can be described by an $O(3)$ invariant
effective field theory \cite{CHIRAL1}-\cite{CHIRAL4}
whose leading part contains two parameters only: $\rho_s$ (spin
stiffness) and $c$ (spin-wave velocity)
\begin{equation}
\label{eff}
{\cal A}^{\rm leading}_{\rm eff} = \frac{\rho_s}{2c} 
\int_0^{c/T}d\tau \int d^2x \, 
 \partial_\mu {\bf R}(\tau,x)\partial_\mu {\bf
R}(\tau,x) \,,\qquad {\bf R}^2=1\,,
\end{equation}
where $\mu=0,1,2$ refer to $\tau,x_1,x_2$, respectively. In their 
work\cite{CHN}, Chakravarty, Halperin and Nelson have used renormalization
group to connect this effective theory with the $d=2$ classical $O(3)$
non-linear $\sigma$-model and, among other results, predicted
the asymptotic, small temperature behavior of the correlation length
$\xi \sim c/2\pi\rho_s \exp(2\pi\rho_s/T)$. The exact mass gap of
the $\sigma$-model\cite{EXACT} and the two-loop corrections lead finally to
the asymptotic expression\cite{HN}
\begin{equation}
\label{xias}
\xi = \frac{e}{8} \frac{c}{2\pi \rho_s} \exp(\frac{2\pi \rho_s}{T})
\left( 1-\frac{T}{4\pi \rho_s}+ O(T^2)\right) \,.
\end{equation}
It has been demonstrated in a careful numerical study\cite{BBGW} that
the explicitly given terms in eq.~(\ref{xias}) are consistent
with the numerical data
at very large correlation lengths (low temperatures) for
$S=1/2$. At moderate correlation lengths, 
however experiments\cite{S12EXP}-\cite{S52EXP3}
numerical data\cite{BBGW}-\cite{NUMUNP}, series expansions
\cite{SERIE} and a semiclassical
model\cite{PHEN} indicate a significant discrepancy which is increasing
rapidly with $S$. It was subsequently 
realized\cite{CROSS1,CROSS2,SERIE} that the quantum
Heisenberg model contains a sequence of crossovers (depending on $T$ and
the parameters $J$ and $S$), and eq.~(\ref{xias}) is valid only in a 
corner called the region of 'renormalized classical scaling'.

This situation is unsatisfactory since the basic feature of
the quantum Heisenberg model in eq.~(\ref{heis}), namely that it can be
mapped onto a simpler model, is valid beyond the region of renormalized
classical scaling. The purpose of this paper is to determine the
corrections to eq.~(\ref{xias}) making it applicable at moderate
correlation lengths also and for any value of $S$ including the large
$S$ limit ('classical scaling region').

\begin{figure}[htb]
\begin{center}
\leavevmode
\epsfxsize=110mm
\epsfbox{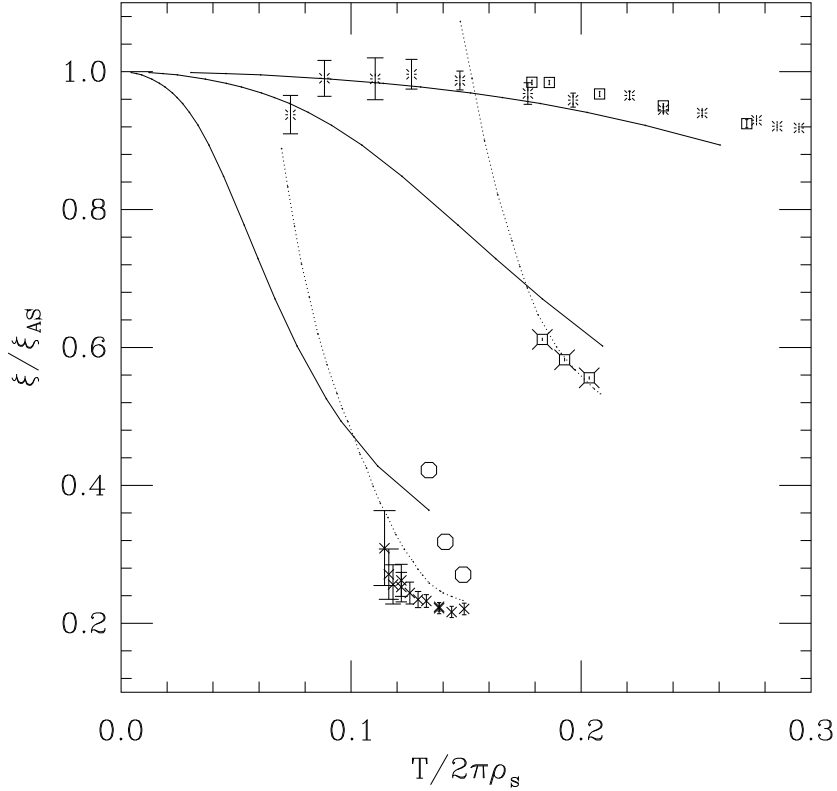}
\end{center}
\caption{The corrected theoretical prediction for the
correlation length normalized by the asymptotic behavior (including the
$O(T)$ corrections) in
eq.~(\ref{xias}) for spin 1/2, 1 and 5/2 (the
upper, middle and lower solid lines, respectively). Correlation lengths
larger than 10 lattice units are considered only.
The MC data are from
ref.\cite{NUM12} ($S=1/2$), ref.\cite{NUM10} ($S=1)$, the 
experimental points
at $S=5/2$ are taken from ref.\cite{S52EXP3}, while the circles come from
series expansion\cite{SERIE}. The dotted lines correspond to a
semiclassical model \cite{PHEN} which performs well at small correlation
lengths but does not have the correct asymptotic behavior.For $\xi_{AS}$,
at $S=1/2$ the values $\rho_s= 0.180$ and $c=1.657$\cite{BBGW}, 
while at $S=1$ and $5/2$ the SWE results\cite{SWE} were used.
Similarly, $\gamma$ was connected  
to $T/2\pi\rho_s$ with the help of the SWE results.}
\label{fig}
\end{figure}

We calculated the corrections to eq.~(\ref{xias}) (which are due to
cut-off effects in the quantum Heisenberg model) in leading non-trivial
order of spin-wave expansion and obtained
\begin{equation}
\label{xicorr}
\xi' = \frac{e}{8} \frac{c}{2\pi \rho_s} \exp(\frac{2\pi \rho_s}{T})
\left( 1-\frac{T}{4\pi \rho_s}\right)\exp(-C(\gamma))\,,\qquad
\gamma=\frac{2JS}{T} \,,
\end{equation}
where the correction factor $\exp(-C(\gamma))$ is given in Table~1 for
different $\gamma$ values. Here $\gamma=2JS/T \sim \Lambda^{\rm cut}/T
(1+O(1/S))$, where $\Lambda^{\rm cut}=c/a$ and 
$a$ is the lattice unit. In
the limit $T \rightarrow 0$, ($S$ fixed), we have 
$\gamma \rightarrow \infty$,
and $C \sim \gamma^{-2}$, leading to the old result in
eq.~(\ref{xias}). For large $S$ and $\rho_s \gg T \gg \Lambda^{\rm cut}$
(classical scaling region), we have $\gamma 
\rightarrow 0$ and $C(\gamma)=\pi/2
+ \ln 8+ \ln \gamma + O(\gamma)$ which gives
\begin{equation}
\label{xicl}
\frac{1}{a}\xi_{\rm classical}=\frac{\exp(-\frac{\pi}{2})}{\sqrt{32}}
 \,\frac{e}{8} \frac{T}{2\pi \rho_s} \exp(\frac{2\pi \rho_s}{T})
\left( 1-O(\frac{\Lambda^{\rm cut}}{T})\right)\,,
\end{equation}
where $\rho_s \sim JS^2 \sim \rho_{class}$. The fact that the prefactor
$\exp(-\pi/2)/\sqrt{32}$ comes out automatically from the corrections is
a non-trivial test on our result.\footnote{Eq.~(\ref{xicl}) has 
the expected form
\cite{CROSS2}: for $S \rightarrow \infty$,  the quantum 
Heisenberg model goes
over to the standard lattice regularized 2D $O(3)$ $\sigma$-model with
bare coupling $g_B = JS^2/T$. The prefactor in eq.~(\ref{xicl}) is the
ratio between the renormalization group invariant scales on the lattice
and in the $\overline {MS}$ renormalization scheme calculated 
long time ago in 
ref.\cite{PAR}.} For intermediate
$\gamma$ values, $\exp(-C(\gamma))$ gives a significant factor as shown
in fig.~\ref{fig}, where, for correlation lengths larger than 10,
the corrected theoretical prediction is compared with
MC and series expansion data and with a semiclassical
model.\footnote{Unfortunately, some of the most interesting large
correlation length MC results at $S=1$ and 5/2 \cite{NUMUNP} are not yet
published and missing from this figure.}

We shall first discuss the steps leading to  eq.~(\ref{xicorr}), then we
close the paper with a few remarks.

\begin{table}
\begin{center}
\begin{tabular}{||l|l|l|l|l|l||} \hline
$\gamma$   & $\exp(-C)$ &  $\gamma$   & $\exp(-C)$ 
 &  $\gamma$   & $\exp(-C)$   \\ \hline
0.0125 & 2.1718 &1.2 & 0.4280 & 4.5 & 0.9410   \\
0.025 & 1.1341 &1.4 & 0.4934 & 5.0 & 0.9538 \\
0.05 & 0.6180 &1.5 & 0.5256 & 5.5 & 0.9623 \\
0.1 & 0.3658 &1.75 & 0.6021 & 6.0 & 0.9691 \\
0.2 & 0.2528 &2.0 & 0.6704   & 7.0 & 0.97771\\
0.4 & 0.2288 &2.25  & 0.7288  & 8.0 & 0.98311 \\ 
0.5 & 0.2399 &2.5 & 0.7775  & 10.0 & 0.98931 \\
0.6 & 0.2576 & 3.0 & 0.8484 & 15.0 & 0.99529  \\
0.8 & 0.3055 &3.5 & 0.8936 & 20.0 & 0.99736  \\
1.0 & 0.3640 & 4.0 & 0.9223 & 30.0 & 0.99883  \\ \hline
\end{tabular}
\end{center}
\caption{The correcting factor $\exp(-C)$ as the function $\gamma=2JS/T$.}
\label{tabla1}
\end{table}

\noindent
2D {\it quantum Heisenberg model vs. the classical} 2D $O(3)$ 
{\it non-linear} $\sigma$-{\it model}: In order to extend the result in 
eq.~(\ref{xias}) we recall that the quantum spin model can be mapped
onto the 2D $\sigma$-model
\begin{equation}
\label{sigma}
{\cal A}_\sigma = \frac{1}{2g} \int d^2x \,\partial_i {\bf e}(x)
\partial_i {\bf e}(x) \,,
\qquad {\bf e}^2=1\,, \qquad i=1,2
\end{equation}
under the condition that the correlation length $\xi$ of the quantum
model satisfies $\xi \gg c/T$ and $\xi \gg a$, where $c/T$ is the length
scale defined by the temperature. In
particular, one is not forced to consider the parameter region only
where the leading effective field theory in eq.~(\ref{eff}) correctly
describes the magnon physics. This is important, since eq.~(\ref{eff})
does not give account of the cut-off effects in the quantum Heisenberg
model. The cut-off effects are related to the fact that the non-magnon
length scales like $c/\rho_s$ are not much larger than the lattice unit
$a$. Actually, for large $S$, $c/\rho_s \sim 1/S$ becomes much smaller
than $a$. The cut-off effects enter the effective prescription
eq.~(\ref{eff}) on the 4-derivative level first\cite{CHIRAL4}
and contribute to the
$O(T^2)$ correction in eq.~(\ref{xias}) when $T \rightarrow 0$ with $J$
and $S$ fixed. In this sense eq.~(\ref{eff}) and eq.~(\ref{xias}) become
correct at sufficiently small temperatures for any given $S$. For large
$S$, however this happens only at astronomically large correlation lengths,
while the mapping to the 2D $\sigma$-model is valid much earlier.

Let us turn now to the argument concerning the mapping of the 2D quantum
Heisenberg model onto the classical 2D $\sigma$-model.\footnote{For
related arguments and different wording, see \cite{HN,QLINK}.} 
The partition
function of the quantum Heisenberg model can be represented in terms of
a path integral using coherent states, for example\cite{COH}. The action is
constructed in terms of a 3-component classical field 
${\bf e}(\tau,n_1,n_2)$, ${\bf e}^2=1$, where $n_1,n_2$ are coordinates
of the two-dimensional spatial lattice and $\tau\in (0,c/T)$ is the
continuous coordinate of the periodic imaginary time direction. If the
correlation length $\xi$ is much larger than $c/T$ (which is the case
for small temperatures), in units of $\xi$ we have a thin slab with two
infinite space directions. This is just a special regularization of the
2D non-linear $\sigma$-model in eq.~(\ref{sigma}).

\noindent
{\it Connecting the parameters of the quantum and the classical models}:
The mapping discussed above leads to quantitative predictions if we
connect the parameters of the
quantum Heisenberg model with the coupling $g$ of the non-linear
$\sigma$-model. A possibility is to choose a convenient long-distance
quantity and calculate it in both models. 
The predictions should match leading to the connection
we are looking for.

A convenient low-energy quantity is the free energy density $f$ as
the function of the chemical potential $h$.\footnote{Unlike the magnetic
field, the chemical potential is renormalization group invariant.} This
was the choice in ref.\cite{HN}, where $f(h)-f(0)$ was calculated in
two-loop perturbation theory in the 2D classical $\sigma$-model 
eq.~(\ref{sigma}) and also in the effective model eq.~(\ref{eff}) (after
introducing a chemical potential in both models in an equivalent
way). Matching the results gives\cite{HN}
\begin{equation}
\label{g}
\frac{1}{g(p)}_{|p=T/c}=\frac{\rho_s}{T} + \alpha + \beta
\frac{T}{\rho_s}\,, \qquad \alpha=0\,,
\beta=-\frac{3}{4}\frac{1}{(2\pi)^2}\,,
\end{equation}
where $g(p)$ is the renormalized coupling constant of the $\sigma$-model
at momentum $p$. Using the relation\cite{EXACT}
\begin{equation}
\label{exact}
\xi = \frac{e}{8} \frac{1}{p}
\frac{g(p)}{2\pi}\exp(\frac{2\pi}{g(p)})\left( 1+\frac{1}{8\pi}g(p)+\dots
\right)\,,
\end{equation}
eq.~(\ref{g}) leads to the prediction in eq.~(\ref{xias}).

Calculating $f(h)-f(0)$ in the quantum Heisenberg model with the help of
the effective model in eq.~(\ref{eff}) (rather than using the Hamilton
operator directly) is an elegant, powerful method. The underlying
assumption is that the higher derivative terms in the effective action
${\cal A}_{\rm eff}$, which do not contribute up to the two-loop
order\cite{CHIRAL4}, give controllable, small corrections. As we discussed
before, due to the large cut-off effects at large $S$ in the quantum
model, this assumption is true at very low temperatures (astronomically
large correlation lengths) only.

In order to calculate the corrections to eq.~(\ref{g}) due to cut-off
effects, we have to abandon the effective prescription\footnote{If
$T/\Lambda^{\rm cut}$
is small, the cut-off effects can be taken into
account by including 4-derivative terms 
(with unknown couplings) in eq.~(\ref{eff}).
For $\rho_s \gg T \sim \Lambda^{\rm cut}$, however, the systematic
derivative expansion of chiral 
perturbation theory breaks down.} and work with the Hamilton operator
directly. We shall use spin-wave expansion (SWE) to derive
$f(h)-f(0)$. It is natural to consider $T$ an $O(S)$
quantity when doing thermodynamics in SWE. Then the 3 terms in
eq.~(\ref{g}) are $O(S)\,,O(1)$ and $O(1/S)$. Our SWE runs up to 
second order and so identifies corrections to the first two terms in 
eq.~(\ref{g}).

The chemical potential introduced in the classical $\sigma$-model (see,
eq.~(4) in ref.\cite{HN}) corresponds to an imaginary twist in the
quantum Heisenberg model\footnote{The author is indebted to Uwe Wiese
for explaining the proper way of introducing the chemical
potential into the Heisenberg model.}
\begin{eqnarray}
\label{heish} 
H(h) &=& H(0)+J\sum_n \left\{ (\cosh(ha)-1)
({\bf S}^1_{n+{\hat 1}}{\bf S}^1_n+
{\bf S}^3_{n+{\hat 1}}{\bf S}^3_n) \right. \nonumber \\
& & \left. -i\sinh(ha)({\bf S}^1_{n+{\hat 1}}{\bf S}^3_n-
{\bf S}^3_{n+{\hat 1}}{\bf S}^1_n)\right\} \,.
\end{eqnarray}
Using the Holstein-Primakoff creation and annihilation operators,
expanding $H(h)$ for large $S$, keeping the $O(S^2)$ 
and $O(S)$ terms and
performing a Bogoljubov transformation\footnote{These are standard
manipulations in the literature on SWE\cite{SWE}.} we obtained
\begin{equation}
\label{heisswe}
H(h)= V \epsilon_0(h)+\sum_k \omega_k(h) b_k^\dagger b_k\,,\qquad
[b_k,b_{k'}^\dagger]=\delta_{k,k'}\,,
\end{equation}
where, in the infinite volume limit\footnote{Here and in the following
we suppress powers higher than $h^2$, since those terms in $f(h)$ are
are not universal pieces in the 2D $\sigma$-model.}
\begin{eqnarray}
\epsilon_0(h) &=& -S(S+1)J(d+\frac{1}{2}(ha)^2) +
\frac{1}{2}\int \frac{d^2k}{(2\pi)^2}\omega_k(h) \,,\nonumber \\
\omega_k(h) &=& 2JS[r_k+(ha)^2s_k]^{1/2}\,, \\
r_k &=& \sum_i(1-\cos k_i)\sum_j(1+\cos k_j)\,, \nonumber \\
s_k &=& \frac{1}{2}\cos k_1\sum_i(1-\cos k_i)+d\,,\qquad i,j=1,2
\,.\nonumber
\end{eqnarray}
We denoted by $d$  and $V$ the space dimension ($d=2$ in
our case) and volume, respectively.
The momentum space integral runs over the Brillouin zone.
The simple Hamilton operator in eq.~(\ref{heisswe}) 
describes free magnon
excitations and a zero-point energy. The magnon specie around $k=(0,0)$
picks up a mass $\sim h^2$, the one around $k=(\pi,\pi)$ remains
massless since $s_k$ goes to zero there. It is an easy exercise to show
that the contribution of the zero point energy to the free energy per
unit slab area $f(h)-f(0)$ is $-h^2\rho_s^{SW}/(2T)$, where
$\rho_s^{SW}$ is the SWE result for the spin-stiffness up to and
including $O(S)$. Actually, this should be so in any order of the SWE,
due to the fact that $\rho_s$ is identical to the helicity modulus and
$h$ is an imaginary twist\cite{SIHU,CHIRAL4}. Adding the contribution 
from the magnon
excitations we get:
\begin{equation}
\label{fh1}
f(h)-f(0)=-\frac{1}{2} h^2 \frac{\rho_s^{SW}}{T} +
\frac{1}{a^2}\int \frac{d^2k}{(2\pi)^2} \ln \frac{1-\exp(-\frac{1}{T}
\omega_k(h))}
{1-\exp(-\frac{1}{T}\omega_k(0))}\,.
\end{equation}
We have to locate the $O(h^2)$ part of the integral in
eq.~(\ref{fh1}). The result can be written in the following way:
\begin{equation}
\label{fh2}
f(h)-f(0)=-\frac{h^2}{2}\left \{\frac{\rho_s^{SW}}{T} + \frac{1}{2\pi}[
\ln \frac{h c^{\rm SW}}{T}-\frac{1}{2} - C(\gamma)]\right \}\,,
\end{equation}
where
\begin{eqnarray}
C(\gamma)&=& \frac{\pi}{2}+\ln 8+\ln \gamma \\
& &  + 2\pi \gamma^2\int \frac{d^2k}{(2\pi)^2} \frac{s_k}{\eta_k}
\frac{\eta_k\exp(-\eta_k)-1+\exp(-\eta_k)}{\eta_k[1-\exp(-\eta_k)]}_
{|\eta_k=\gamma(r_k)^{1/2}}\,, \nonumber
\end{eqnarray}
using the notation $\gamma=2JS/T$. For $T \rightarrow 0$, 
$\exp(-C(\gamma))$
behaves like $1-1.05(1)\gamma^{-2}$ and eq.~(\ref{fh2}) 
becomes identical with
the spin-wave expanded form of eq.~(7) in 
ref.\cite{HN} obtained from the
effective theory. (There are additional $\propto (T/\rho_s)^2$
corrections to the $O(T^2)$ term in eq.~(\ref{xias}), of coarse.) 
This is an explicit confirmation of the effective action technique.

The new term $\propto C(\gamma)$ in eq.~(\ref{fh2}) 
modifies the relation
between the parameters giving $\alpha = -C(\gamma)/(2\pi)$ in 
eq.~(\ref{g}). Eq.~(\ref{exact})
leads then to the result quoted in eq.~(\ref{xicorr}).

Let us close this paper with some remarks. It would be interesting
to calculate the next order in the SWE which would 
test the full 2-loop
result for consistency and add subleading cut-off corrections. This
calculation seems to be feasible.

In the large $S$ limit the square lattice quantum Heisenberg model
becomes identical to the standard lattice regularized $O(3)$
$\sigma$-model for {\it any} correlation length. Due to the existence of
powerful cluster MC techniques precise correlation length data are
available from $\xi=O(1)$  up to $O(200)$ (with finite size scaling
techniques even beyond)\cite{O3}
which can be compared with
results in the quantum Heisenberg model from series expansion and MC for
large $S$.

This paper is about cut-off effects in the quantum Heisenberg model. We
should emphasize, this has nothing to do with the cut-off effects in the
2D non-linear $\sigma$-model. We always considered large correlation
lengths and so the cut-off effects in the $\sigma$-model, which are
suppressed as $\sim \xi^{-2}$, are negligible.
    
\noindent
{\bf Acknowledgments}
The author is indebted to Uwe Wiese for explaining how the chemical
potential enters the quantum Heisenberg model in our context. He is also
indebted to Peter Keller-Marxer for making his $S=1$ and 5/2 MC data
available before publication and also for his help in compiling data from
the literature. The author thanks him and Ferenc Niedermayer
for the  critical reading of the manuscript and Rebecca Christianson, 
Norbert Elstner, Rajiv Singh and Ruggero Vaia for providing data in
numerical form for the figure.
While working on this subject, the author enjoyed the warm hospitality
of Shanta deAlwis, Tom DeGrand and Anna Hasenfratz at the Dept. of
Physics, Boulder. This work was partially supported by the US Department
of Energy and by Schweizerischer Nationalfond.

 
\newcommand{\PL}[3]{{Phys. Lett.} {\bf #1} {(19#2)} #3}
\newcommand{\PR}[3]{{Phys. Rev.} {\bf #1} {(19#2)}  #3}
\newcommand{\NP}[3]{{Nucl. Phys.} {\bf #1} {(19#2)} #3}
\newcommand{\PRL}[3]{{Phys. Rev. Lett.} {\bf #1} {(19#2)} #3}
\newcommand{\PREPC}[3]{{Phys. Rep.} {\bf #1} {(19#2)}  #3}
\newcommand{\ZPHYS}[3]{{Z. Phys.} {\bf #1} {(19#2)} #3}
\newcommand{\ANN}[3]{{Ann. Phys. (N.Y.)} {\bf #1} {(19#2)} #3}
\newcommand{\HELV}[3]{{Helv. Phys. Acta} {\bf #1} {(19#2)} #3}
\newcommand{\NC}[3]{{Nuovo Cim.} {\bf #1} {(19#2)} #3}
\newcommand{\CMP}[3]{{Comm. Math. Phys.} {\bf #1} {(19#2)} #3}
\newcommand{\REVMP}[3]{{Rev. Mod. Phys.} {\bf #1} {(19#2)} #3}
\newcommand{\ADD}[3]{{\hspace{.1truecm}}{\bf #1} {(19#2)} #3}
\newcommand{\PA}[3] {{Physica} {\bf #1} {(19#2)} #3}
\newcommand{\JE}[3] {{JETP} {\bf #1} {(19#2)} #3}
\newcommand{\FS}[3] {{Nucl. Phys.} {\bf #1}{[FS#2]} {(19#2)} #3}

\eject


\begin{thebibliography}{99}

\bibitem{S12EXP}
B.~Keimer et al., Phys.Rev.B46, 14034 (1992);
M.~Greven et al., Phys.Rev.Lett.72, 1096 (1994) and Z.Phys.B 96,
465 (1995).

\bibitem{S10EXP}
K.~Nakajima et al., Z.Phys.B96, 479 (1995);
R.~J.~Birgeneau et al., Phys.Rev.B41, 2514 (1990) and references therein.

\bibitem{S52EXP1}
R.~A.~Cowley et al., Phys.Rev.B15, 4292 (1977).

\bibitem{S52EXP2}
Y.~S.~Lee et al., Eur.Phys.J B5, 15 (1998).

\bibitem{S52EXP3}
R.~L.~Leheny et al., cond-mat/9809178, to appear in Phys.Rev.Lett.82 (1999).

\bibitem{CHIRAL1}
S.~Weinberg, Phys.Rev.Lett.17, 616 (1966);
J.~Gasser and H.~Leutwyler, Ann.Phys.(NY)158, 142 (1984) and
Nucl.Phys.B250, 465 (1985).

\bibitem{CHIRAL2}
M.~E.~Fisher and V.~Privman, Phys.Rev.B32, 447 (1985);
E.~Brezin and J.~Zinn-Justin, Nucl.Phys.B257 [FS14], 867 (1985).

\bibitem{CHIRAL3}
H.~Neuberger and T.~Ziman, Phys.Rev.B39, 2608 (1989).

\bibitem{CHIRAL4}
P.~Hasenfratz and H.~Leutwyler, Nucl.Phys.B343, 241 (1990);
P.~Hasenfratz and F.~Niedermayer, Z.Phys.B92, 91 (1993).

\bibitem{CHN}
S.~Chakravarty, B.~I.~Halperin and D.~R.~Nelson, Phys.Rev.B39, 
2344 (1989).

\bibitem{EXACT}
P.~Hasenfratz, M.~Maggiore and F.~Niedermayer, Phys.Lett.B245, 
522 (1990);
P.~Hasenfratz, and F.~Niedermayer, Phys.Lett.B245, 529 (1990).

\bibitem{HN}
P.~Hasenfratz, and F.~Niedermayer, Phys.Lett.B268, 231 (1991).

\bibitem{BBGW}
B.~B.Beard, R.~J.~Birgeneau, M.~Greven and U.~-J.~Wiese,
Phys.Rev.Lett.80, 1742 (1998).

\bibitem{NUM12}
M.~Greven, U.~-J.~Wiese and R.~J.~Birgeneau, Phys.Rev.Lett.77, 
1865 (1996);
J.-K.~Kim D.~P.~Landau and M.~Troyer, Phys.Rev.Lett.79, 1583 (1997);
J.-K.~Kim  and M.~Troyer, Phys.Rev.Lett.80, 2705 (1998).

\bibitem{NUM10}
K.~Harada, M.~Troyer and N.~Kawashima, J.~Phys.Soc.Jap.67, 
1130 (1998).

\bibitem{NUMUNP}
B.~B.~Beard, V.~Chudnovsky, P.~Keller-Marxer and U.~-J.~Wiese, 
in preparation.

\bibitem{SERIE}
N.~Elstner et al., Phys.Rev.Lett.75, 938 (1995).

\bibitem{PHEN}
A.~Cuccoli et al. J.Phys.Cond.Mat.7, 7891 (1995);
A.~Cuccoli, V.~Tognetti,  R.~Varia and P.~Verruchi, 
Phys.Rev.Lett.77, 3439 (1996) and
Phys.Rev.B56, 14456 (1997).

\bibitem{CROSS1}
A.~V.~Chubukov and S.~Sachdev, Phys.Rev.Lett.71, 169 (1993);
A.~V.~Chubukov, S.~Sachdev and J.~Ye, Phys.Rev.B49, 11919 (1994);
N.~Elstner, R.~L.~Glenister, R.~R.~P.~Singh and A.~Sokol, 
Phys.Rev.B51, 8984
(1995).

\bibitem{CROSS2}
A.~Sokol, N.~Elstner and R.~R.~P.~Singh, cond-mat/9505148.

\bibitem{PAR}
G.~Parisi, Phys.Lett.B92, 133 (1980).

\bibitem{QLINK} S.~Chandrasekharan and U.-J.~Wiese, hep-lat/9609042.

\bibitem{COH}
J.~R.~Klauder and B.-S.~Skagerstam: Coherent States, World
Scientific, 1985.

\bibitem{SWE}
T.~Oguchi, Phys.Rev. 117, 117 (1960);
J.~Igarashi, Phys.Rev.B46, 10763 (1992);
C.~J.~Hamer, Z.~Weihong and J.~Oitmaa, Phys.Rev.B50, 6877 (1994).

\bibitem{SIHU}
R.~R.~P.~Singh and D.~A.~Huse, Phys.Rev.B40, 7247 (1989).

\bibitem{O3}
U.~Wolff, Nucl.Phys.B334, 581 (1990);
B.~Alles, G.~Cella, M.~Dilaver and Y.~G\"und\"uc, 
hep-lat/9808003 and
references therein.




\end{thebibliography}
\end{document}